\documentclass[12pt, a4paper]{article}
\usepackage{amsfonts, amssymb}
\usepackage{graphicx}
\usepackage{latexsym,amsmath,color,array}

\usepackage{enumerate}
\usepackage{bbding}
\usepackage{amssymb}
\usepackage{amsmath}
\usepackage{graphicx}
\usepackage{amsmath}
\usepackage{bbm}
\usepackage{mathtools}
 \usepackage{color}
 \usepackage{array}
 \usepackage{subfigure}
 \usepackage{multirow}
 \usepackage[dvipsnames]{xcolor}
 \usepackage{makecell}
 \usepackage{colortbl}
\def\1{\mathbb{I}}

\newcounter{thm}[section]
\newcounter{appen}[section]
\newcounter{assum}[section]

\begin{document}

\title{A modelling methodology for social interaction experiments}
\author{Susan C. Fennell\footnote{MACSI, Department of Mathematics and Statistics, University of Limerick, Limerick, Ireland} \hspace{1cm}
James P. Gleeson\footnotemark[1] \hspace{1cm}
Michael Quayle\footnote{Department of Psychology, University of Limerick, Limerick, Ireland} \footnotemark[3]\\
Kevin Durrheim\footnote{Department of Psychology, University of KwaZulu-Natal, Pietermaritzburg, KwaZulu-Natal, South Africa} \hspace{1cm}
Kevin Burke\footnotemark[1]
 }
\date{}

\maketitle

\begin{abstract}
Analysis of temporal network data arising from online interactive social experiments is not possible with standard statistical methods because the assumptions of these models, such as independence of observations, are not satisfied.  In this paper, we outline a modelling methodology for such experiments where, as an example, we analyse data collected using the Virtual Interaction Application (VIAPPL) --- a software platform for conducting experiments that reveal how social norms and identities emerge through social interaction.  We apply our model to show that ingroup favouritism and reciprocity are present in the experiments, and to quantify the strengthening of  these behaviours over time. Our method enables us to identify participants whose behaviour is markedly different from the norm. We use the method to provide a visualisation of the data that highlights the level of ingroup favouritism, the strong reciprocal relationships, and the  different behaviour of participants in the game. While our methodology was developed with VIAPPL in mind, its usage extends to any type of social interaction data.

\smallskip


\end{abstract}

\newpage

\section{Introduction}
Social interaction is a driving force in shaping people's opinions, beliefs, and behaviours. Understanding how the emergence and evolution of social structures, such as norms and identities, depends on social interaction is an important avenue of research in social psychology, sociology, economics and related disciplines \cite{Drury2000,Drury2009,Reicher1995,Spears2015,Postmes2005,Ioannides2012}.
Empirical studies have shown that norms emerge through interaction, however these studies either lack experimental control \cite{Haslam2007}, or they cannot analyse how the individual interactions produce the emergent norms \cite{Ruscher2006,Postmes2000}. Typically, empirical studies produce relatively limited amounts of data, and, therefore, an important aspect of such studies is the usage of statistical models to infer results. Of particular interest in this context is the nature of participant-to-participant and group-level interactions, which produce temporal network data.

VIAPPL, the Virtual Interaction APPLication, is a software platform for running social experiments in a controlled setting allowing researchers to study how social structures emerge though social interaction (see www.viappl.org).
Participants, or players, in the experiments are avatars in a virtual game where they exchange tokens with the other players over a number of rounds. VIAPPL games can be flexibly defined to manipulate various features of social interaction, and here we describe a specific game that will be analyzed in this paper. In this game players only communicate through token exchange (not, for example, face-to-face or though a chat function on the platform), and it is this token exchange which provides a measure of social interaction.
At the end of each round, players are presented with a network diagram of all individual player-to-player token exchanges from that round; the diagram also indicates group membership (since players are randomly assigned to one of two groups at the beginning of the experiment). Players use this information to inform their decision making in the next round, and, in this way, the interactions between players influence behaviour and shape norms in a given game. 

The data from a VIAPPL experiment is temporal network data. The observations corresponding to one player (i.e., who this player exchanged tokens with over all rounds) are not likely to be independent of each other, nor to those of other players: they are interconnected since the players, exposed to the same game history, consider their own previous actions and the actions of others when deciding on their next move. Because of this interconnectivity, the assumptions underlying many standard statistical procedures used in analysing more classical (static, non-network) data (e.g., t-test, ANOVA, or normal linear regression \cite{Blanca2018, Nizam2013}), are not satisfied by VIAPPL data; thus, a new approach is required. 

VIAPPL data has been analysed previously to investigate the evolution of ingroup favouritism under different conditions, such as inequality \cite{Durrheim2016}. However, the data in that study was aggregated, and, so, analysis at the level of individual interactions was not possible. By contrast in this paper, we propose a modelling framework which, among other things, incorporates this level of detail. This is achieved by linking token exchanges between pairs of individuals in a given round through a regression model. Unlike standard procedures, we do not assume a null distribution for the error terms (normal or otherwise), but, rather, make use of a null model for player behaviour. The sequence of round-based regression coefficients maps out a trajectory over time, enabling us to see how norms become more or less important as the game progresses. Furthermore, we provide a method for detecting players who have very different behaviour from the other players in a game. Note that the proposed methodology is applicable beyond the VIAPPL setting to other kinds of social interaction data.

The outline of the paper is as follows. First, we describe the VIAPPL game environment in detail before discussing our modelling approach. We then apply the model to investigate the presence of ingroup favouritism and reciprocity, and provide a method for visualising the data based on the model. Finally, we conclude by discussing the results of the analysis and further applications for the modelling framework. 

\section{VIAPPL game environment}
\begin{figure}
    \includegraphics[width=\textwidth]{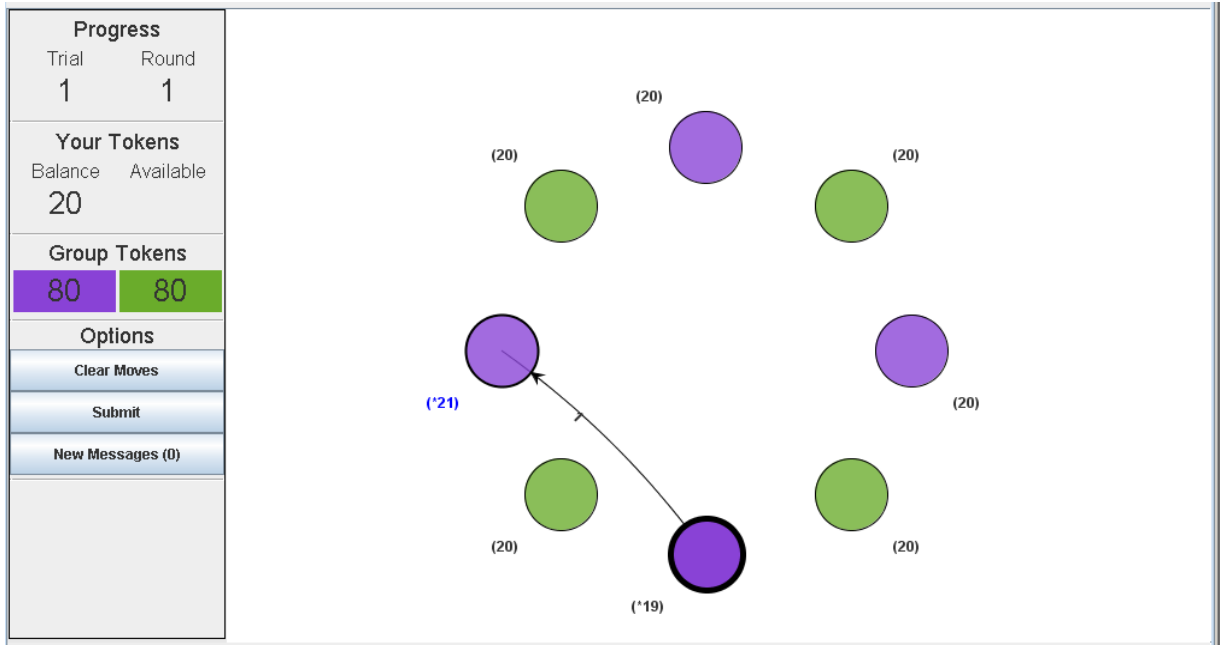}
   \caption{{\bf VIAPPL screen presented to a player as they select who they will allocate a token to.}}
    \label{fig1}
\end{figure}

VIAPPL, the Virtual Interaction APPlication, is a software platform for conducting experiments in social psychology. Participants of the experiments are avatars in a game-like environment and are referred to as players. They observe other players as nodes in a network, as shown in Fig \ref{fig1}, and they interact by exchanging tokens over a number of rounds. At the start of the game, players are randomly allocated to one of two groups, and group membership is highlighted by the node colour. The number of tokens each player has is indicated next to their node, with group totals shown to the left of the screen; note that, in the game shown in Fig \ref{fig1}, all players start with 20 tokens. A player recognises their own node by the thick black border, and selects a node to allocate a token (where they may select their own node); each player allocates one token per round. The ego player in Fig \ref{fig1}, who is in the purple group, has given a token to another player in the purple group (i.e., ingroup giving), as indicated by the arrow. The token balance of the ego player is reduced by one, while the balance of the other player increases by one.

\begin{figure}
    \includegraphics[width=\textwidth]{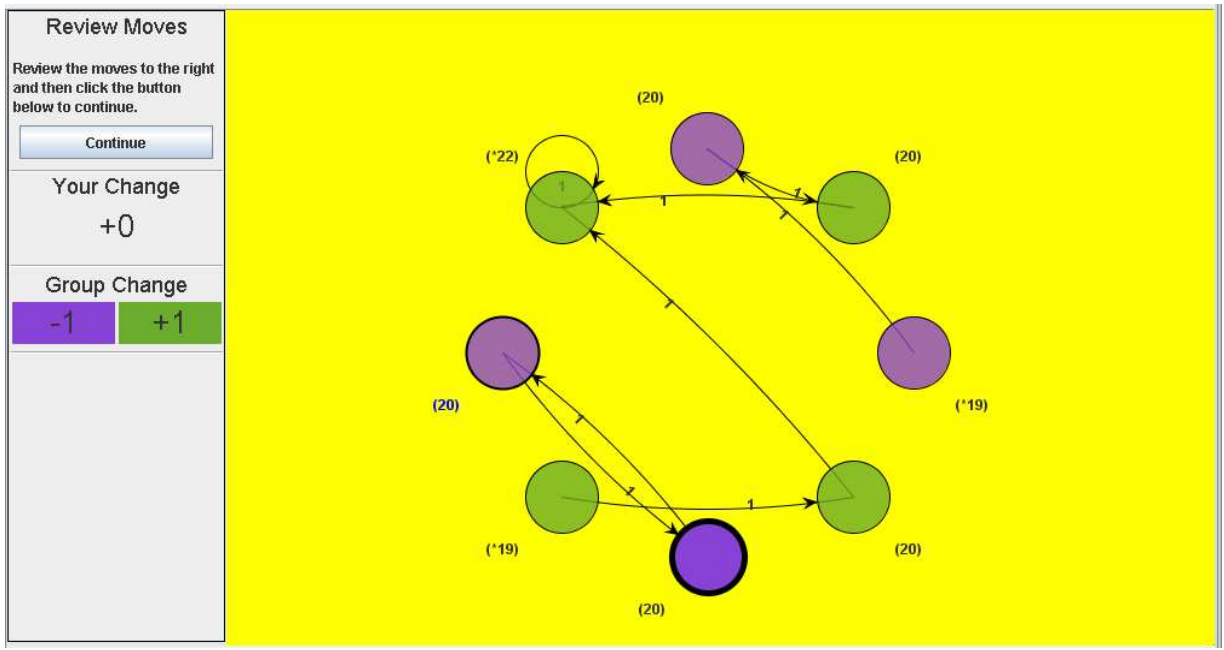}
    \caption{{\bf VIAPPL screen presented to players at the end of a round, displaying all token allocations from that round.}}
    \label{fig2}
\end{figure}

Once all players have made their move, a new screen appears which displays all token allocations from that round in a directed network diagram, as in Fig \ref{fig2}. Note that in this instance the ego player has received a token from the player they gave their token to. An example of self-giving can be seen by the green node at the top left of the screen, as indicated by the self-directed arrow. Each player reviews the interactions of the round before the next round commences.

Many of the features in the VIAPPL platform can be changed to accommodate different experimental setups. The initial number of tokens assigned to each player need not be homogeneous, and may differ with respect to the group. The number of players, groups, and players per group can also be varied. Nodes may be located anywhere on the screen, not necessarily in a circular format. Nodes can also be different shapes or images, allowing players to express individuality so that they may be defined beyond their group membership. In this paper, we apply our methodology to a balanced setup like that of Fig \ref{fig1} where there are two groups, an equal number of players per group, and each player starts with the same number of tokens; however, the methodology we present can be used in the other setups.

\section{Modelling approach}
As described in the introduction, individual interactions are key to the formation and recreation of norms, and, thus, we model this data by considering the token exchanges between pairs of individuals as the variable of interest. More specifically, we choose the response variable $Y_{ijt}$ to be the number of tokens player $i$ has {\it received} from player $j$ up to round $t$. There are two fundamental norms that may occur in a VIAPPL game: reciprocation, where players give more often to players they have received tokens from, and ingroup favouritism, where players prefer to give to players in their own group. Reciprocation can be measured by investigating the effect of $Y_{jit}$, the number of tokens player $i$ has {\it given} to player $j$ up to round $t$, on the response $Y_{ijt}$. 
Ingroup favouritism can be measured by investigating the effect of $G_{ij}$, a binary variable indicating whether $i$ and $j$ are in the same group ($G_{ij}=0$) or not ($G_{ij}=1$), on the response. Combining these two effects, we propose the linear model,
\begin{equation}
Y_{ijt} = \alpha + \rho\,Y_{jit} + \gamma\,G_{ij} + \epsilon_{ijt},
\label{eqn1}
\end{equation}
where $\alpha$ is an intercept, $\rho$ is the reciprocity effect, and $\gamma$ is the group effect, while $\epsilon_{ijt}$ is an error term. Of course, we may fit this model at several time points, $t$, and, indeed, we do this in our analysis. Thus, we could highlight the $t$ dependence in the coefficients via $\alpha_t$, $\rho_t$, and $\gamma_t$, respectively; however, we suppress that dependence here for notational convenience. Note that, informally, we could write the model as
\begin{align*}
&\text{``Tokens received from an individual''} \\
&\qquad= \alpha + \rho\,\text{``Tokens given to that individual''} + \gamma\,\text{``Do groups differ?''}.
\end{align*}

As was mentioned in the previous section, VIAPPL players have the option to give tokens to themselves. However, the concepts of reciprocity and ingroup favouritism do not apply in this case. As such, the $Y_{iit}$ terms need to be treated differently to the $Y_{ijt}$ ($i\neq j$) terms. The two cases --- self-giving and non-self-giving --- can be handled by an extended model,
\begin{equation*}
Y_{ijt} = (\alpha + \psi S_{ij}) + (\rho + \rho^*S_{ij}) Y_{jit} + \gamma\, G_{ij} + \epsilon_{ijt},
\end{equation*}
where $S_{ij}$ is a binary variable denoting the self such that $S_{ij}=1$ when $i=j$ and $S_{ij}=0$ otherwise. (Note that there is no $G_{ij}S_{ij}$ term since this is always zero.)
The $i\ne j$ case of course returns to Eq (\ref{eqn1}), whereas, the $i=j$ case becomes
\begin{equation*}
Y_{iit} = (\alpha + \psi) + (\rho + \rho^*) Y_{iit} +  \epsilon_{iit}
\end{equation*}
since $S_{iit}=1$ and $G_{iit}=0$. Clearly, $\psi = -\alpha$ and $\rho^* =1-\rho$ to yield $Y_{iit} = Y_{iit}$ (with $\epsilon_{iit}=0$), and, so, these new coefficients are wholly determined by the other coefficients in the model. Therefore, it is sufficient to consider only the $i\ne j$ cases, and, importantly, as we have just demonstrated, there is no loss of information in doing so. In practice, this is equivalent to removing the $i=j$ elements from the dataset and fitting the model of Eq (\ref{eqn1}); writing this formally, our model is
\begin{equation}
Y_{ijt} = \alpha + \rho\,Y_{jit} + \gamma\,G_{ij} + \epsilon_{ijt}, \qquad i \ne j.
\label{eqn2}
\end{equation}

\subsection{Assessing significance of effects}
To assess the significance of a particular effect ($\rho$ or $\gamma$), we may specify a null hypothesis, e.g., $H_0: \gamma = 0$ for the group effect, and,  then, consider the extent to which the estimate from experimental data, $\hat\gamma$, differs from zero. Typically, in a linear model, the assumptions of normality and independence of errors are made, which, in turn, impose a t-distribution on the (standardised) estimated effect under the null hypothesis \cite{Nizam2013}; this result permits straightforward significance testing.

\begin{figure}[!h]
\centering
\includegraphics{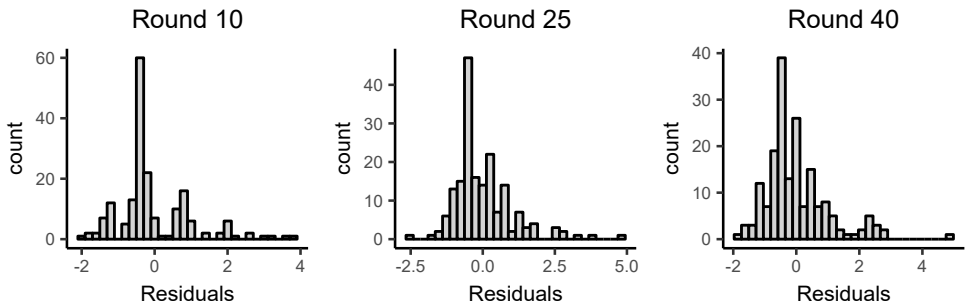}
\caption{{\bf Histograms of the errors from the model of Eq (\ref{eqn2}) at {\boldmath$t \in \{10, 25, 40\}$} for game 1 of the data analysed in the Results section.} There we analyse three other games, but the histograms shown here are representative of what we see in all games.}
\label{fig3}
\end{figure}

In our setting, standard assumptions are violated due to constraints within the experimental setup (e.g., if players give one token per round, then $\sum_{i=1}^n Y_{ijt} = t$, where $n$ is the number of players), and the high interconnectivity of such social interaction data. Indeed, Fig \ref{fig3} shows errors from the model of Eq (\ref{eqn2}) fitted to experimental data at $t \in \{10,25,40\}$, and it is clear that these are highly non-normal in all cases, but particularly in earlier rounds. Therefore, instead of specifying a null hypothesis for model coefficients in conjunction with normally distributed errors, we specify a null {\it model} for the underlying player behaviour. This null model approach is quite common in network analysis \cite{Wandelt2019}, and has been used in the analysis of online influence and opinion evolution \cite{Kerckhove2016}, but is not so prevalent in social psychology; interestingly, the approach has also been used in a range of ecology applications \cite{Farine2017,Gotelli1996}.

We make use of a null model in which players give at random, i.e., a player has an equal probability of choosing each player in the network to give their token to in a given round. We simulate a game according to this null model using an agent-based model \cite{Jackson2017,Bonabeau2002,Conte2013}, and fit the linear model of Eq (\ref{eqn2}) to this simulated game. This procedure yields a random draw of estimated coefficients for a null game, say, $(\tilde\alpha,\tilde\rho,\tilde\gamma)$, and, repeating this a large number of times (we use 10,000 replicates for very high accuracy), provides the distribution for each coefficient under the null behaviour. By comparing estimated coefficients from the real data to these simulated distributions, we can assess the significance of each effect. Note that, although we use the ``giving at random'' null model, it being analogous to the hypothesis of ``no effect'' used commonly in statistics, the real data could be compared to other behaviours of potential interest, e.g., a null model in which players give twice as often to their ingroup as they do to the outgroup.

\subsection{More general models}

A small extension of Eq (\ref{eqn2}), and one which we consider in our analysis, includes an interaction between $Y_{jit}$ and $G_{ij}$ via
\begin{equation}
Y_{ijt} = \alpha + \rho\,Y_{jit} + \gamma\,G_{ij} + \delta\,G_{ij}\,Y_{jit} + \epsilon_{ijt}, \qquad i \ne j,
\label{eqn3}
\end{equation}
where $\delta$ adjusts the reciprocity effect in light of the group status, and vice versa.

Another extension, but not one we consider here, would be the inclusion of additional predictor variables via
\begin{equation}
Y_{ijt} = \alpha + \rho\,Y_{jit} + \gamma\,G_{ij} + \beta\cdot X_{ijt} + \epsilon_{ijt}, \qquad i \ne j,
\label{eqn4}
\end{equation}
\noindent
where $X_{ijt}$ is a vector of predictors, $\beta$ is a corresponding vector of regression coefficients, and ``$\cdot$'' is the dot product. These additional predictors could relate to the individual, such as age or gender; to aspects of the game's history, such as the number of tokens $j$ received from $i$ over some particular window of time, $Y_{ji{t_2}}-Y_{ji{t_1}}$; or to conditions of the experimental setup (if results are pooled from multiple setups), such as whether or not individuals started with the same number of tokens. Of course, Eqs \ref{eqn3} and \ref{eqn4} could be combined by considering both $G_{ij}\,Y_{jit}$ and $ X_{ijt}$, and, furthermore, other interactions could be considered, i.e., $X_{ijt}$ terms with $Y_{jit}$ and $G_{ij}$, and $X_{ijt}$ terms with each other.

Note that, in the above, we have chosen the response variable $Y_{ijt}$ to be the number of tokens that player $i$ received from player $j$ {\it up to} round $t$. However, other choices could be made, for example, $Y_{ijt}$ could be a binary variable indicating whether or not player $i$ received a token from player $j$ {\it in} round $t$; in this case, a binary regression model (e.g., logistic or probit) would be more appropriate than a linear model, and, more generally, one could consider other generalised linear models \cite{McCullagh1989}.

\section{Results}
To demonstrate our methodology, we consider data collected from 4 VIAPPL experiments, each of which had a different group of 14 participants (56 participants in total). The experiments were carried out in the University of KwaZulu-Natal, South Africa. All participants provided written informed consent to participate in the study, which had been approved by the Human Sciences Research Ethics Committee of the University of KwaZulu-Natal. The games each had 2 groups of 7 players, and each player started with 40 tokens. At the start of the game, players were instructed to give one token per round and were told that the game would last for 40 rounds. Since players started with 40 tokens, they could not run out of tokens at any point in the game. (These games differ slightly from the example presented in Figs \ref{fig1} and \ref{fig2} in that there are more players and tokens.)

\subsection{Exploratory analysis}

Before applying the proposed model to the data, we first carry out some  exploratory analysis to gain some initial insight. Here, we retain the $ii$ data points (self-giving), but, as described above, they are removed when applying the model.

\begin{figure}
\centering
\includegraphics{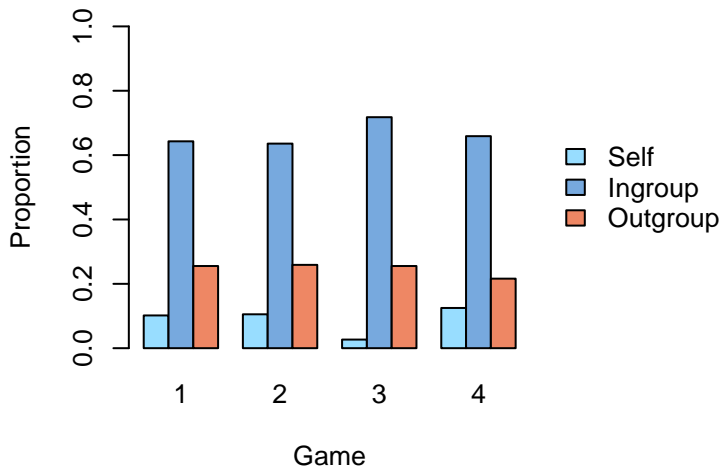}
\caption{{\bf Proportion of tokens received from ingroup, outgroup and self in each game.}}
\label{fig4}
\end{figure}

\begin{figure}[!h]
\begin{center}
\includegraphics{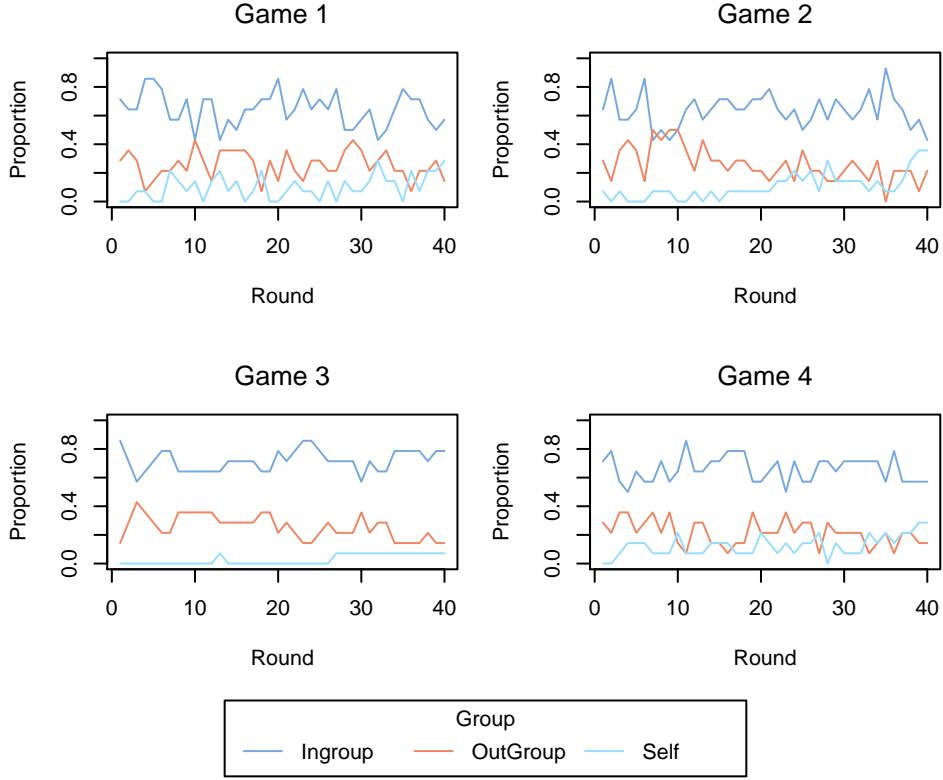}
\end{center}
\caption{{\bf Proportion of tokens received from the ingroup, outgroup and self at each round.}}
\label{fig5}
\end{figure}

Fig \ref{fig4} shows the proportion of tokens players received from their ingroup, the outgroup and from themselves over the course of the game. Interestingly, players receive at least twice as many tokens on average from their ingroup as they do from their outgroup (and recall that group membership is randomly assigned). The values of each of the three proportions are remarkably similar across games, albeit game 3 has less self-giving and more ingroup giving. Fig \ref{fig5} displays these proportions calculated at each round. In almost all cases, players receive more tokens from their ingroup than their outgroup. The amount of self-giving increases slightly as time goes on in all games, and, in fact, it increases beyond the outgroup proportion towards the end of games 1, 2, and 4.

\begin{figure}[!h]
\begin{center}
\includegraphics{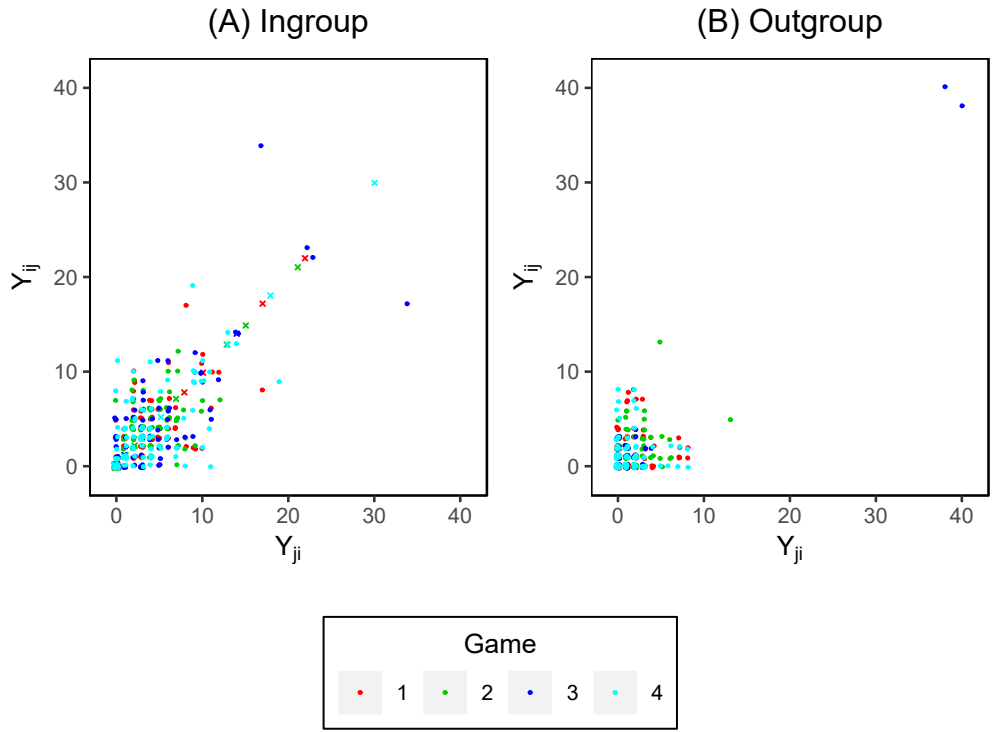}
\end{center}
\caption{{\bf The number of tokens received is plotted against the number of tokens given for (A) players in the same group and (B) players in different groups.} The $\times$ symbol indicates self-giving. The points at $(38,40)$ and $(40,38)$ in plot (B) correspond to two players who reciprocated for almost the entirety of the game.}
\label{fig6}
\end{figure}

The scatterplots in Fig \ref{fig6} show the relationship between the number of tokens received and the number of tokens given over the whole course of the game, split by group. Each point $({Y}_{ij},{Y}_{ji})$ corresponds to a pair of players $(i,j)$, indicating that player $i$ {\it received} ${Y}_{ij}$ tokens from player $j$ and {\it gave} ${Y}_{ji}$ tokens to player $j$. When referring to the whole game we avoid writing $Y_{ij,40}$ for convenience.
The scatterplots are symmetric about the diagonal ${Y}_{ij}={Y}_{ji}$, since, for each pair of players $(i,j)$, there are two points $({Y}_{ij},{Y}_{ji})$ and $({Y}_{ji},{Y}_{ij})$.
In particular, cases of self-giving corresponds to the point (${Y}_{ii},{Y}_{ii}$) which, necessarily, appears on the diagonal in Fig \ref{fig6}A (as it is an ingroup exchange). Points tend to be closer to the diagonal in the case of ingroup giving (Fig \ref{fig6}A), indicating that players reciprocate with members of their own group. The plot corresponding to the outgroup, Fig \ref{fig6}B, shows points further away from the diagonal, indicating less reciprocation in general.

Table \ref{table1} displays the correlation coefficient between tokens received, ${Y}_{ij}$, and tokens given, ${Y}_{ji}$, split by the ingroup and outgroup, and by game; the closer this value is to one, the greater the reciprocity. In all games, there is a reasonably strong ingroup reciprocity, whereas outgroup reciprocity is much weaker in games 1, 2, and 4. On the other hand, in game 3 the outgroup reciprocity is very strong. However, this high correlation coefficient is mainly driven by two players with unusual behaviour, and when their token exchanges are removed it reduces to 0.08. These two players, from different groups, formed a reciprocal relationship in which they exchanged tokens in almost all of the rounds; this is quite unlike what we observe in other players across all games. Their token exchanges are clearly visible in the top-right corner of Fig \ref{fig6}B, lying far away from all other points (even when compared to the ingroup setting of Fig \ref{fig6}A); later in our analysis, these two players are also identified as being highly unusual based on the model.

\begin{table}[!ht]
\centering
\caption{
{\bf Correlation between ${Y}_{ij}$ and ${Y}_{ji}$ for the ingroup and the outgroup.}}
\begin{tabular}{ccccc}
  \hline
  & \multicolumn{4}{c}{ \bf Game}\\ \hline
 & 1 & 2 & 3 & 4 \\ 
  \hline
  Ingroup & 0.68 & 0.69 & 0.80 & 0.73 \\ 
  Outgroup & 0.09 & 0.33 & 0.97 & -0.06 \\ 
   \hline
\end{tabular}
\label{table1}
\end{table}

\subsection{Model for the full game}
 
We apply the linear model, described in the Modelling Approach section, to the full game, i.e., exchanges up to, and including, the final round ($t=40$). The response, $Y_{ij}$, is the number of tokens player $i$ received from player $j$ over the course of the game. We consider four models which are special cases of Eq (\ref{eqn3}):
\begin{enumerate}[(i)]
    \item reciprocity effect only ($\gamma = \delta = 0$),
    \item group effect only ($\rho = \delta = 0$),
    \item additive reciprocity and group effects ($\delta=0$), and
    \item interacting reciprocity and group effects.
\end{enumerate}
Table \ref{table2} displays the $R^2$ values for these four models in each of the four games. Note that the highest $R^2$ values are seen in game 3, but this is driven by the two anomalous players mentioned earlier. Looking at the other games, we see that the models with additive effects have larger $R^2$ values than the models with only one of the two effects, whereas the inclusion of the interaction effect increases the $R^2$ to a much lesser extent.

\begin{table}[!ht]
\centering
\caption{
{\bf $R^2$ for models for the full game.}}
\begin{tabular}{cccccc}
  \hline
&& \multicolumn{4}{c}{ \bf Game}\\ \hline
\multicolumn{2}{c}{ \bf Model} & 1 & 2 & 3 & 4 \\ 
  \hline
Reciprocity &  $Y_{ji}$ & 0.22 & 0.25 & 0.78 & 0.26 \\ 
Group       &  $G_{ij}$ & 0.24 & 0.29 & 0.08 & 0.23 \\ 
Additive    &  $G_{ij}+Y_{ji}$ & 0.31 & 0.35 & 0.78 & 0.33 \\ 
Interaction &  $G_{ij}*Y_{ji}$ & 0.33 & 0.35 & 0.79 & 0.37 \\ 
   \hline
\end{tabular}
\label{table2}
\end{table}

\begin{figure}
\centering
\includegraphics{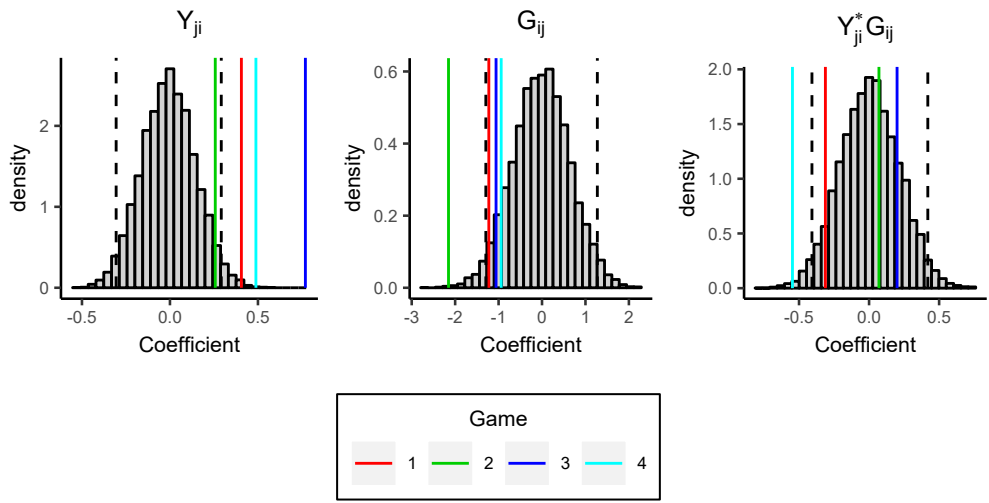}
\caption{{\bf Coefficients for the interaction effects models.} Game coefficients are the vertical coloured lines. The histograms show the model coefficients from the simulated (null model) games. $95\%$ of values lie between the black dashed lines.}
\label{fig7}
\end{figure}

\begin{figure}
\centering
\includegraphics{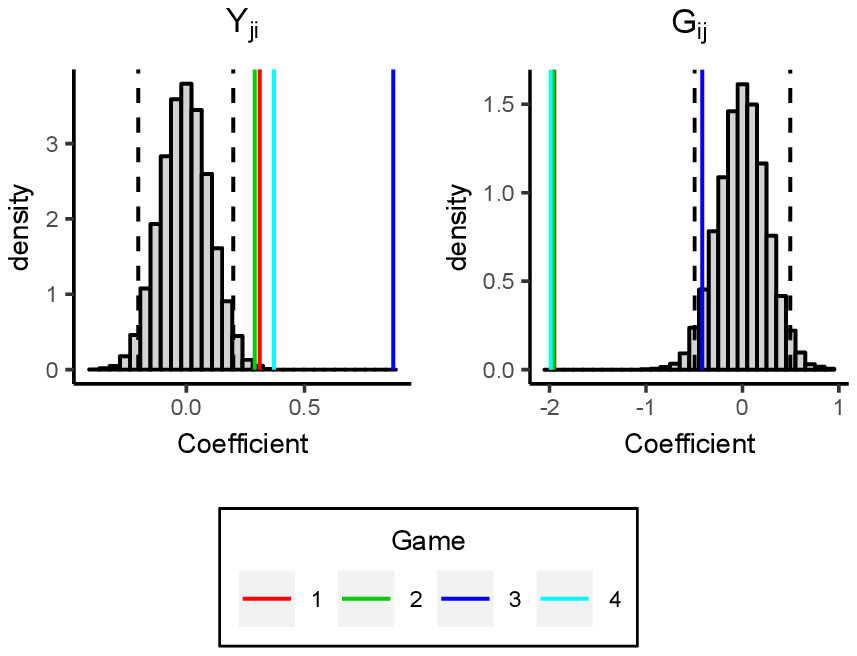}
\caption{{\bf Coefficients for the additive effects models.} Game coefficients are the vertical coloured lines. The histograms show the model coefficients from the simulated (null model) games. $95\%$ of values lie between the black dashed lines.}
\label{fig8}
\end{figure}

Fig \ref{fig7} displays the coefficients for the interaction effects models, along with histograms of the coefficients obtained from the simulated (null model) data. Note that the interaction term is within the $95\%$ bounds for games 1, 2, and, 3, and only just outside of these bounds for game 4, i.e., the interaction effect is not strong as was also suggested by Table \ref{table2}. We therefore consider the additive effects models in more detail; Fig \ref{fig8} displays the coefficients against the reference null distributions, and Table \ref{table3} provides the numeric values of these coefficients along with p-values computed based on the null distribution. It is clear the group effect is strong and highly significant statistically, with players in games 1, 2, and 4 receiving approximately two fewer tokens on average from outgroup members than ingroup members. The reciprocity effect is not as strong, but it is still highly significant for all four games; its positive value indicates that players receive more tokens from players they give tokens to. Again it is clear that the behaviour in game 3 differs from the others, with a group effect which is non-significant (at the $5\%$ level), and a much stronger reciprocity effect. 

\begin{table}[!ht]
\centering
\caption{{\bf Coefficients for the additive effects models.} P-values are in brackets next to the coefficients.}
\begin{tabular}{ccccc}
  \hline
& \multicolumn{4}{c}{ \bf Game}\\ \hline
 & {\bf 1} & {\bf 2} & {\bf 3} & {\bf 4} \\ 
  \hline
{\bf Intercept} & 2.960 (0.787) & 3.010 (0.656) & 0.600 (\textless 0.001) & 2.770 (0.757) \\ 
  $\mathbf{Y_{ji}}$ & 0.310 (0.001) & 0.290 (0.004) & 0.870 (\textless 0.001) & 0.370 (\textless 0.001) \\ 
  $\mathbf{G_{ij}}$ & -1.950 (\textless 0.001) & -1.960 (\textless 0.001) & -0.420 (0.106) & -1.990 (\textless 0.001) \\ 
   \hline
\end{tabular} 
\label{table3}
\end{table}

\subsection{Model for each round}

\begin{figure}[!h]
\centering
\includegraphics{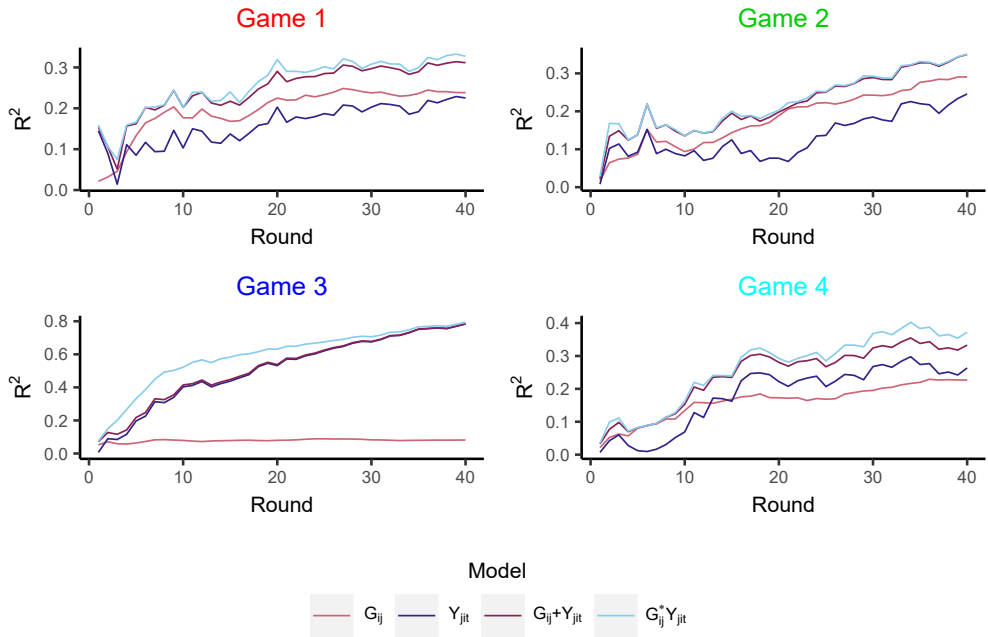}
\caption{{\bf $R^2$ for each model at each round.}}
\label{fig9}
\end{figure}

While the previous section focused on the exchanges over the whole game, i.e., $t=40$, we now fit the various models, but at all time points $t\in\{1,2,\ldots,40\}$ to analyse the evolution of the effects over time. Fig \ref{fig9} displays the $R^2$ for each model at each round. We see that the $R^2$ values are generally increasing over time, suggesting that the behaviour becomes more predictable, perhaps as players settle into some normative behaviour. It is also clear that, over essentially all time points, the additive effects models improve on the single reciprocity and group effects models, while the interaction effects models are only slightly better than the additive effects models; the exception to this is game 3 (the unusual game) in the earlier rounds where the interaction effects models do appear to improve the fit (albeit this is not the case in the latter rounds).

\begin{figure}[!h]
\centering
\includegraphics{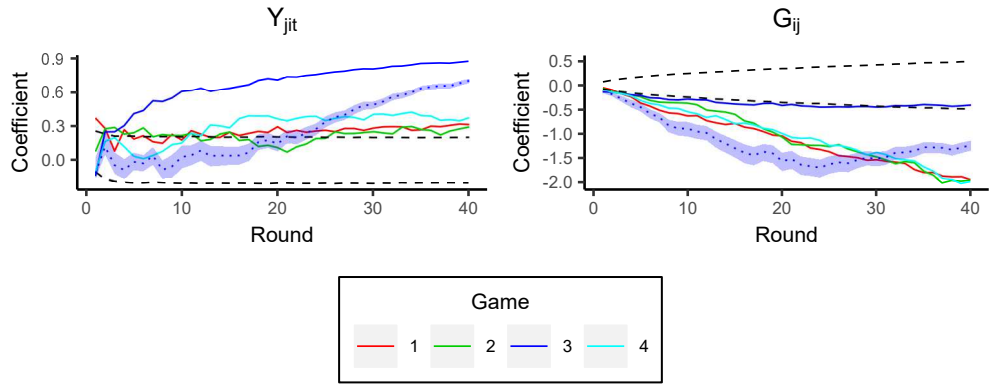}
\caption{{\bf Coefficients for the additive effects models at each round}. Solid coloured lines are the game coefficients. Black dashed lines are the $95\%$ confidence bands from the simulated data. The dotted line is the average coefficient from the simulated data for game 3 (explained in the Game 3 section), and the surrounding shaded region displays the $95\%$ bounds.}
\label{fig10}
\end{figure}

The coefficients for the additive effects models over each round are shown in Fig \ref{fig10}. For games 1, 2, and 4, the group effects are significant at almost all rounds, i.e., the norm of giving to the ingroup is apparent from the outset. Moreover, since the group effects get further away from the $95\%$ bounds with time, this behaviour strengthens with time. By contrast, the reciprocity effect is non-significant in the earlier rounds, i.e., it takes longer for individuals to build reciprocal links. Indeed, recall from Fig \ref{fig6} that individuals reciprocate more with individuals from the ingroup. This, in combination with the fact that the group effect is established much earlier in the game, perhaps suggests that the group membership provides a context for reciprocal links to develop. The coefficients, and their trajectories over time, are remarkably similar across these three games, particularly the group effect, i.e., there is repeatability in the dynamics over different groups of participants; of course, game 3 is quite different from the other games with a much larger reciprocity effect, and much smaller (non-significant) group effect (which we consider in more detail in the next section).

\subsection{Game 3}

Up to now we have not explicitly referred to the identity of a specific player in the experimental data, beyond pointing out that an unusual pair of interactions exists in game 3 (which was clearly visible in Fig \ref{fig6}B). Of course, there are 14 individuals in each game, and, therefore, the player index $i$ lies in the set $\{1,\ldots,14\}$. The numbering of players is completely arbitrary, but, for example, we use the player's position in the network diagram, starting at the top node (labelled ``player 1'') and moving in a clockwise direction. Labelling in this way, in game 3, player 3 gave 38 (out of a possible 40) tokens to player 8, while player 8 gave 40 tokens to player 3. This level of reciprocity was far higher than between any other pairs, both in game 3 and the other games. In this section, we investigate the extent to which these two players contributed to the large reciprocity effect for that game, and, indeed, the small group effect (note that these players are in different groups).

To examine the influence of players 3 and 8 on the estimated model coefficients, we create semi-simulated data in which these players are replaced with null players who give at random. This data was created by artificially replaying the game, keeping all token exchanges as per the observed data, apart from the tokens given by players 3 and 8 which are instead allocated at random. This has the effect of altering only the values of $Y_{i,3,t}$ and $Y_{i,8,t}$, i.e., in the real data, $Y_{i,8,t}$ equals $t$ for $i=3$ and zero otherwise, whereas, in the semi-simulated data, $Y_{i,8,t}$ equals $t/14$ on average $\forall i$; similarly, in the semi-simulated data, $Y_{i,3,t}$ equals $t/14$ on average $\forall i$.

In total, we generated 10,000 semi-simulated datasets, and applied the additive effects model to each. This yields a distribution for the model coefficients with the unusual reciprocal behaviour of players 3 and 8 removed. The average coefficients from this distribution along with $95\%$ bounds are shown in Fig \ref{fig10} by the dotted line and shaded area. Indeed, compared to the real game 3, the reciprocity effect is reduced, but still becomes larger than that of the other games; this suggests that an unusual level of reciprocity exists in game 3 even having removed the influence of players 3 and 8. Note that, in Fig \ref{fig6}A, there are points in positions (17,34) and (34,17) corresponding to players 1 and 5, and points in positions  (23,22) and (22,23) corresponding to players 2 and 12; they are much nearer to the top right corner than any other (non-self-giving) points, indicating sustained reciprocity. These two reciprocal pairs are not as unusual as the player 3 and player 8 pairing, since they correspond to players in the same group, and, as we have seen, ingroup giving is normative behaviour. For this reason, the group effect in the semi-simulated game 3 is still very strong. However, it does reduce in later rounds as the reciprocity effect becomes stronger.

\subsection{Influence metric}

Players 3 and 8 in game 3 were initially identified as being unusual through the exploratory analysis, i.e., they are clear outliers in Fig \ref{fig6}B. However, building on the approach used in the previous section, we now develop a more systematic, model-based approach for identifying anomalies by way of their influence on the model coefficients. A standard statistical technique for identifying influential observations is to remove an observation from the dataset, refit the model to this reduced dataset, and compute the difference between the new and original coefficients, so-called ``dfbetas'' \cite{Cook1982}; large changes in the coefficients signify influential observations. 

In our context of social interaction data, however, simple removal of observations changes the structure of the data due to the interrelatedness of these observations. For example, removing $Y_{ji}$ leads to player $i$ having given $40 - Y_{ji}$ tokens, whereas all others give 40 (without the loss of generality, we mean $t=40$ here and avoid the $t$ subscript for convenience.); removing the set of all tokens given by player $i$, $(Y_{1,i},\ldots,Y_{14,i})$, creates a non-player who still received tokens; removing player $i$ entirely creates imbalance in the group sizes, and, furthermore, all tokens related to this individual (giving and receiving) become unaccounted for. These removals alter the data structure,  which is why, in examining the joint influence of players 3 and 8 in game 3, we did not simply {\it remove} these players but, rather, {\it replaced} them with simulated players. This maintains the structure of the data where token exchanges are redistributed. Thus, in order to determine the influence of a single player on the model coefficients, we use that approach here but with an individual player, rather than a pair.

As in the previous section, we create a semi-simulated game where everything remains as per the observed data apart from the actions of player $i$ who is replaced by a null player who gives at random. For the $k^{th}$ semi-simulated game we fit the model to obtain the altered coefficients $\hat\rho_{k,t}^{-i}$ and $\hat\gamma_{k,t}^{-i}$ for $t=1,\ldots40$, where the superscript ``$-i$'' indicates that the influence of player $i$ is removed. These coefficients map out a curve which could be compared to those of Fig \ref{fig10}. However, it is not practical to view 14 altered coefficient curves for both coefficients across four games. We therefore compute the distances between altered and original curves using an $\ell_1$ norm for each simulation,
\begin{equation*}
\lVert\hat\rho-\hat\rho_{k}^{-i}\rVert_{1}, \quad\qquad \lVert\hat\gamma-\hat\gamma_{k}^{-i}\rVert_{1},
\end{equation*}
where $\hat\rho = (\hat\rho_{1},\ldots,\hat\rho_{40})$ and $\hat\gamma = (\hat\gamma_{1},\ldots,\hat\gamma_{40})$ are the vectors of original coefficients, and $\hat\rho_k^{-i} = (\hat\rho_{k,1}^{-i},\ldots,\hat\rho_{k,40}^{-i})$ and $\hat\gamma_k^{-i} = (\hat\gamma_{k,1}^{-i},\ldots,\hat\gamma_{k,40}^{-i})$ are the vectors of altered coefficients for simulation $k$. Then, we average these over simulations to obtain the influence metrics,
\begin{equation}
d\rho_{i} = \frac{1}{N}\sum_{k=1}^{N}\lVert\hat\rho-\hat\rho_{k}^{-i}\rVert_{1}, \quad\qquad d\gamma_{i} = \frac{1}{N}\sum_{k=1}^{N}\lVert\hat\gamma-\hat\gamma_{k}^{-i}\rVert_{1},
\end{equation}
where $N$ is the number of simulations; we use $N=10{,}000$. In order to make these metrics more comparable, we standardise them by dividing by the average value across all players in the game,
\begin{equation}
\frac{d\rho_{i}}{\sum_{i=1}^{14} d\rho_{i} / 14}, \quad\qquad \frac{d\gamma_{i}}{\sum_{i=1}^{14} d\gamma_{i} / 14}.
\end{equation}
Values greater than 1 indicate player $i$ is more influential than the average player.

These values are shown in Table \ref{table4}, where we also highlight values greater than 2 as these correspond to players whose influence is more than double the average. While there are players with influence scores greater than 2 in all games (only just in some cases), there are also players with scores much larger than this. In game 3, players 3 and 8 have influence scores greater than 5 for both variables indicating that their behaviour strongly influenced the estimated model coefficients; this is in line with our findings in the previous section. These players were far more influential than players in any of the other games. Interestingly, there are two players in game 4 with high influence scores. Player 6 has a large influence score for the group effect. It turns out that this player gave no tokens to outgroup players, and, although ingroup giving is normative, it is unusual to have avoided the outgroup entirely (on average in game 4, players gave 8.6 tokens to outgroup players). Player 7 has large influence scores for both effects, albeit more for the reciprocity effect than the group effect. This player reciprocated 35 tokens throughout the game, of which 34 were with ingroup players (on average in game 4, players reciprocated 20.6 tokens).

\begin{table}[ht]
\centering
\caption{{\bf Standardised influence metrics.} Values larger than 2 are marked with a * to indicate players with more influence on the model.}
\begin{tabular}{c|ll|ll|ll|ll}
  \hline 
 \multicolumn{1}{c|}{\multirow{3}{*}{{\bf Player}}} & \multicolumn{8}{c}{{\bf Game}} \\\cline{2-9}
 \multicolumn{1}{c|}{} & \multicolumn{2}{c}{1} & \multicolumn{2}{c}{2}& \multicolumn{2}{c}{3}& \multicolumn{2}{c}{4}\\ \cline{2-9}
 & $Y_{ji}$ & $G_{ij}$ & $Y_{ji}$ & $G_{ij}$ & $Y_{ji}$ & $G_{ij}$ & $Y_{ji}$ & $G_{ij}$ \\ 
  \hline
1 & 0.28  & 0.26$\phantom{*}$  & 0.69  & 0.41$\phantom{*}$  & 0.73  & 0.61  & 0.39  & 0.27  \\ 
  2 & 0.91  & 1.30  & 0.59  & 0.62  & 0.50  & 0.42  & 1.63  & 0.43  \\ 
  3 & 2.01* & 0.44  & 0.51  & 0.35  & 5.19* & 5.44* & 1.52  & 1.13  \\ 
  4 & 1.39  & 0.70  & 0.46  & 1.51  & 0.33  & 0.16  & 0.53  & 1.18  \\ 
  5 & 0.43  & 0.80  & 0.42  & 1.21  & 0.39  & 0.59  & 0.58  & 0.21  \\ 
  6 & 0.91  & 1.65  & 1.22  & 1.04  & 0.24  & 0.18  & 0.69  & 3.71* \\ 
  7 & 1.38  & 0.89  & 1.62  & 0.94  & 0.18  & 0.11  & 3.76* & 2.72* \\ 
  8 & 1.10  & 0.32  & 1.05  & 1.48  & 5.09* & 5.46* & 0.27  & 0.21  \\ 
  9 & 0.44  & 0.85  & 1.72  & 0.85  & 0.15  & 0.21  & 0.37  & 0.33  \\ 
  10 & 0.25  & 0.80  & 1.23  & 1.68  & 0.16  & 0.16  & 0.84  & 1.27  \\ 
  11 & 0.66  & 0.85  & 0.61  & 0.63  & 0.11  & 0.07  & 1.53  & 0.84  \\ 
  12 & 2.52* & 1.96  & 2.23* & 0.80  & 0.56  & 0.34  & 0.42  & 0.63  \\ 
  13 & 0.59  & 1.85  & 0.96  & 1.69  & 0.09  & 0.05  & 0.37  & 0.32  \\ 
  14 & 1.11  & 1.32  & 0.69  & 0.79  & 0.28  & 0.21  & 1.10  & 0.74  \\ 
   \hline
\end{tabular}
\label{table4}
\end{table}

\subsection{Visualising the data}

\begin{figure}[!h]
\centering
\includegraphics{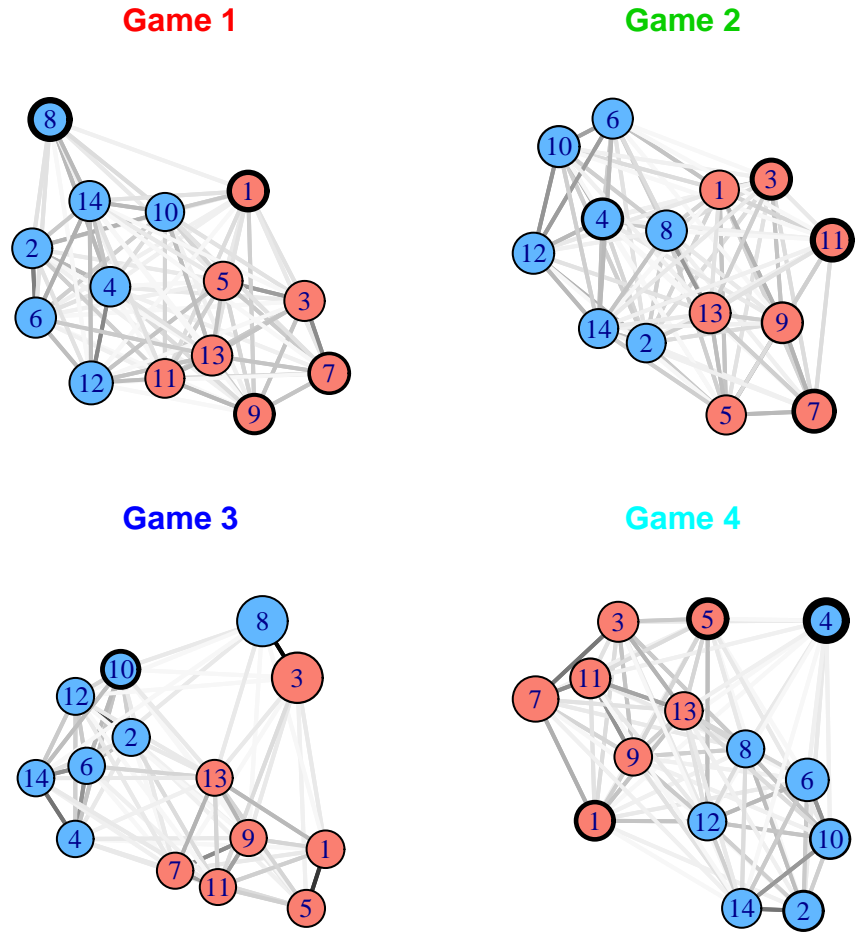}
\caption{{\bf The network of players for each game.} The node colour indicates the group while the node size relates to size of the influence scores from Table \ref{table4} (larger nodes indicate higher scores). Edges between nodes are drawn if players exchanged tokens, where the edge colour corresponds to the total number of exchanges between the players in either direction (darker edges correspond to more exchanges). The size of the border around a node indicates the level of self-giving for that player. }
\label{fig11}
\end{figure}

Fig \ref{fig10} visualises the game dynamics over time, aggregated over individuals via model coefficients. We now provide a complementary network diagram visualisation in Fig \ref{fig11} which provides an alternative view in terms of how individual player's contribute to the aggregate normative behaviour. This is a weighted network produced based on the total number of token exchanges between pairs of players (in either direction) by the end of the game (albeit such diagrams could be produced at any time point in the game).

From the network diagram, we get a sense of which specific players reciprocated with each other, the level of ingroup or outgroup giving, and how unusual players are based on their influence score. The network layout was generated using the Fruchterman-Reingold layout algorithm in the \texttt{igraph} package in R \cite{Fruchterman1991,Csardi2006}, a force-directed algorithm which locates nodes more closely together if their edge weight is larger.

 As we would expect from our earlier analysis, there is a clear split between the two groups, showing the tendency of players to interact more with members of their ingroup. Players who are located more centrally in the network interacted with both groups, e.g., players 10 and 11 in game 1, whereas players at the periphery of the network interacted more exclusively with their ingroup, e.g., players 2 and 3 in game 1. Of course, players 3 and 8 in game 3 stand out clearly, both due to their node size (influence score), and the fact that they are they are located further away from other nodes. This is due to the fact that they essentially only interacted with each other, however there are still weak links with other players meaning that, throughout the game, other players did interact with them to some extent. 
Player 4 in game 4 is also quite distinctive. This player self-gave tokens 30 times during the game (visible in the thick border), and, much like players 3 and 8, is located further away from other nodes due to the lower level of interaction with other players.

\section{Discussion}

VIAPPL, the Virtual Interaction APPLication, is a software platform for running experiments to study how social norms and identities emerge through social interaction. VIAPPL data has been analysed previously, but in an aggregated form which did not previously allow the effect of individual interactions to be studied \cite{Durrheim2016}. We have introduced a modelling framework that includes this individual level detail. Our models indicate that ingroup favouritism is a prominent normative behaviour which emerges right from the outset of the game; this is in agreement with \cite{Durrheim2016}. We have additionally found that players  have a tendency to reciprocate, and this behaviour becomes more pronounced as the game progresses, i.e., it takes time for a reciprocal relationship to develop, whereas ingroup favouritism is more immediate. The ability to detect reciprocation is only possible by modelling the interactions between individuals as we have done in this paper.

Interestingly, the estimated group and reciprocity effects are quite similar across 3 of the 4 games. However, one of the games produced quite different results from the rest. We showed that this was mainly due to two players from different groups who reciprocated with each other for most rounds of the game (albeit there were other high-reciprocating players in that game). Our model-based influence metric identified these individuals as being unusual, i.e., their behaviour (high reciprocation with an outgroup member) was much different from the average behaviour in that game. When these two players were removed from the data the results of the model were much more similar to results from the other games.
Our visualisation of the network of players complements the output from the linear model, as it indicates the contribution of individual players to the overall dynamics estimated by the model.

We note that our suggested influence metric (for detecting individuals differing from the norm) could be used to detect anomalous individuals in various settings. One topical example of this type of application is bot detection on Twitter, where the focus is on determining user accounts that display unusual behavioural patterns \cite{Ferrara2016}. It would be interesting to apply the approach to such data to determine whether or not bot accounts are indeed identified, and visualise the network as per Fig.~\ref{fig11}; such information could complement other methods currently employed for bot identification \cite{Varol2017}.

While developed for VIAPPL data, this methodology can be used for any kind of social interaction data, and, of course, these interactions need not be ``token exchange'' (this is just the social interaction mechanism within the VIAPPL platform.) Our proposed approach does not impose distributional assumptions on the error terms, nor independence of observations, which is useful when these standard modelling assumptions are not met. The use of a null model makes this approach particularly suited to scenarios where the data is produced according to some rule-based process, for example, in iterated prisoner's dilemma experiments in game theory \cite{Axelrod1987}. However, this does not prevent its use in other settings once a suitable agent-based model (see \cite{Jackson2017,Bonabeau2002,Conte2013}) is defined for the scenario under study.

\section{Acknowledgments}
The authors acknowledge funding from the Irish Research Council (SCF), Science Foundation Ireland (JPG, grant numbers 16/IA/4470 and 16/RC/3918) and the European Research Council (ERC) under the European Union's Horizon 2020 research and innovation programme (MQ, grant agreement No. 802421).

\bibliographystyle{plain}
\bibliography{VIAPPL}

\begin{thebibliography}{10}

\bibitem{Axelrod1987}
Robert Axelrod et~al.
\newblock The evolution of strategies in the iterated prisoner’s dilemma.
\newblock {\em The dynamics of norms}, pages 1--16, 1987.

\bibitem{Blanca2018}
Mar{\'{\i}}a~J. Blanca, Rafael Alarc{\'{o}}n, and Roser Bono.
\newblock Current practices in data analysis procedures in psychology: What has
  changed?
\newblock {\em Frontiers in Psychology}, 9, dec 2018.

\bibitem{Bonabeau2002}
E.~Bonabeau.
\newblock Agent-based modeling: Methods and techniques for simulating human
  systems.
\newblock {\em Proceedings of the National Academy of Sciences}, 99(Supplement
  3):7280--7287, may 2002.

\bibitem{Conte2013}
Rosaria Conte, Rainer Hegselmann, and Pietro Terna.
\newblock {\em Simulating social phenomena}, volume 456.
\newblock Springer Science \& Business Media, 2013.

\bibitem{Cook1982}
R.~Dennis. Cook and Sanford Weisberg.
\newblock Residuals and influence in regression.
\newblock {\em Monographs on statistics and applied probability}, 1982.

\bibitem{Csardi2006}
Gabor Csardi and Tamas Nepusz.
\newblock The igraph software package for complex network research.
\newblock {\em InterJournal}, Complex Systems:1695, 2006.

\bibitem{Drury2000}
John Drury and Steve Reicher.
\newblock Collective action and psychological change: The emergence of new
  social identities.
\newblock {\em British journal of social psychology}, 39(4):579--604, 2000.

\bibitem{Drury2009}
John Drury and Steve Reicher.
\newblock Collective psychological empowerment as a model of social change:
  Researching crowds and power.
\newblock {\em Journal of Social Issues}, 65(4):707--725, 2009.

\bibitem{Durrheim2016}
Kevin Durrheim, Michael Quayle, Colin~G. Tredoux, Kim Titlestad, and Larry
  Tooke.
\newblock Investigating the evolution of ingroup favoritism using a minimal
  group interaction paradigm: The effects of inter- and intragroup
  interdependence.
\newblock {\em {PLOS} {ONE}}, 11(11):e0165974, nov 2016.

\bibitem{Farine2017}
Damien~R. Farine.
\newblock A guide to null models for animal social network analysis.
\newblock {\em Methods in Ecology and Evolution}, 8(10):1309--1320, apr 2017.

\bibitem{Ferrara2016}
Emilio Ferrara, Onur Varol, Clayton Davis, Filippo Menczer, and Alessandro
  Flammini.
\newblock The rise of social bots.
\newblock {\em Communications of the ACM}, 59(7):96--104, 2016.

\bibitem{Fruchterman1991}
Thomas~MJ Fruchterman and Edward~M Reingold.
\newblock Graph drawing by force-directed placement.
\newblock {\em Software: Practice and experience}, 21(11):1129--1164, 1991.

\bibitem{Gotelli1996}
Nicholas~J Gotelli and Gary~R Graves.
\newblock {\em Null models in ecology}.
\newblock Smithsonian Institution Scholarly Press, 1996.

\bibitem{Haslam2007}
S~Alexander Haslam and Stephen Reicher.
\newblock Identity entrepreneurship and the consequences of identity failure:
  The dynamics of leadership in the bbc prison study.
\newblock {\em Social psychology quarterly}, 70(2):125--147, 2007.

\bibitem{Ioannides2012}
Yannis~M. Ioannides.
\newblock {\em From Neighborhoods to Nations: The Economics of Social
  Interactions}.
\newblock Princeton University Press, 2012.

\bibitem{Jackson2017}
Joshua~Conrad Jackson, David Rand, Kevin Lewis, Michael~I. Norton, and Kurt
  Gray.
\newblock Agent-based modeling: a guide for social psychologists.
\newblock {\em Social Psychological and Personality Science}, 8(4):387--395,
  mar 2017.

\bibitem{Kerckhove2016}
Corentin~Vande Kerckhove, Samuel Martin, Pascal Gend, Peter~J. Rentfrow,
  Julien~M. Hendrickx, and Vincent~D. Blondel.
\newblock Modelling influence and opinion evolution in online collective
  behaviour.
\newblock {\em {PLOS} {ONE}}, 11(6):e0157685, jun 2016.

\bibitem{McCullagh1989}
P.~McCullagh and J.A. Nelder.
\newblock {\em Generalized Linear Models, Second Edition}.
\newblock Chapman \& Hall/CRC Monographs on Statistics \& Applied Probability.
  Chapman \& Hall, 1989.

\bibitem{Nizam2013}
Azhar Nizam, David~G. Kleinbaum, Lawrence~L. Kupper, and Eli Rosenberg.
\newblock {\em Applied Regression Analysis and Other Multivariable Methods}.
\newblock Cengage Learning, Inc, 2013.

\bibitem{Postmes2005}
Tom Postmes, S~Alexander Haslam, and Roderick~I Swaab.
\newblock Social influence in small groups: An interactive model of social
  identity formation.
\newblock {\em European review of social psychology}, 16(1):1--42, 2005.

\bibitem{Postmes2000}
Tom Postmes, Russell Spears, and Martin Lea.
\newblock The formation of group norms in computer-mediated communication.
\newblock {\em Human communication research}, 26(3):341--371, 2000.

\bibitem{Reicher1995}
Stephen~D Reicher, Russell Spears, and Tom Postmes.
\newblock A social identity model of deindividuation phenomena.
\newblock {\em European review of social psychology}, 6(1):161--198, 1995.

\bibitem{Ruscher2006}
Janet~B Ruscher and Elliott~D Hammer.
\newblock The development of shared stereotypic impressions in conversation: An
  emerging model, methods, and extensions to cross-group settings.
\newblock {\em Journal of Language and Social Psychology}, 25(3):221--243,
  2006.

\bibitem{Spears2015}
Russell Spears and Tom Postmes.
\newblock Group identity, social influence, and collective action online.
\newblock {\em The Handbook of the Psychology of Communication Technology, John
  Wiley \& Sons, Oxford}, pages 23--46, 2015.

\bibitem{Varol2017}
Onur Varol, Emilio Ferrara, Clayton~A Davis, Filippo Menczer, and Alessandro
  Flammini.
\newblock Online human-bot interactions: Detection, estimation, and
  characterization.
\newblock In {\em Eleventh international AAAI conference on web and social
  media}, 2017.

\bibitem{Wandelt2019}
Sebastian Wandelt, Xiaoqian Sun, Ernestina Menasalvas, Alejandro
  Rodr{\'{\i}}guez-Gonz{\'{a}}lez, and Massimiliano Zanin.
\newblock On the use of random graphs as null model of large connected
  networks.
\newblock {\em Chaos, Solitons {\&} Fractals}, 119:318--325, feb 2019.

\end{thebibliography}

\end{document}